\documentclass[aps,print,showpacs,twocolumn,10pt]{revtex4}
\usepackage{amssymb}
\usepackage{amsmath}
\usepackage{graphicx}

\input{tcilatex}
\begin{document}

\title{Soliton Solution for the Spin Current in Ferromagnetic Nanowire}
\author{Zai-Dong Li$^{1}$, Qiu-Yan Li$^{1}$, Lu Li$^{2}$, W. M. Liu$^{3}$}
\affiliation{$^{1}$Department of Applied Physics, Hebei University of Technology, Tianjin
300401, China\\
$^{2}$College of Physics and Electronics Engineering, Shanxi University,
Taiyuan, 030006, China\\
$^{3}$Beijing National Laboratory for Condensed Matter Physics, Institute of
Physics, Chinese Academy of Sciences, Beijing 100080, China}

\begin{abstract}
We investigate the interaction of a periodic solution and one-soliton
solution for the spin-polarized current in a uniaxial ferromagnetic
nanowire. The amplitude and wave number of the periodic solution for the
spin current have the different contribution to the width, velocity, and the
amplitude of soliton solution, respectively. Moreover, we found that the
soliton can be trapped only in space with a proper condition. At last we
analyze the modulation instability and discuss dark solitary wave
propagation for the spin current on the background of a periodic solution.
In some special cases the solution can be expressed by the linear
combination of periodic solution and soliton solution.
\end{abstract}

\pacs{05.45.Yv, 75.40.Gb}
\maketitle

\section{Introduction}

The study of magnetoelectronics has received considerable interest for its
technological potential application. Both theoretical and experimental
investigations mainly concentrated on giant magnetoresistance are of
fundamental importance in the understanding of magnetism and applied
interest in the fabrication of magnetic devices. In metallic ferromagnets,
the differences between electronic bands and scattering cross-sections of
impurities for majority and minority spins at the Fermi energy cause
spin-dependent mobilities \cite{Arne}. The difference between spin-up and
spin-down electric currents is called a spin-current. It is a tensor, with a
direction of flow and a spin-polarization parallel to the equilibrium
magnetization vector $\mathbf{M}\left( \mathbf{r},t\right) $ as \cite{S.
Zhang04}%
\begin{equation}
\mathcal{J}=\frac{P\mu _{B}}{eM_{s}}\mathbf{j}_{e}\otimes \mathbf{M}\left( 
\mathbf{r},t\right) ,  \label{spincurrent0}
\end{equation}%
where $P$ is the spin polarization of the current, $\mu _{B}$ is the Bohr
magneton, and $e$ is the magnitude of electron charge. The vector $\mathbf{j}%
_{e}$ tracks the direction of the charge current, $\mathbf{M}\left( \mathbf{r%
},t\right) $ describes the direction of the spin polarization of the
current, and $M_{s}$ is the saturation magnetization.

When the magnetization directions in the systems are not collinear, the
polarization directions of the non-equilibrium accumulations and currents
are not parallel or antiparallel with the magnetizations. This gives rise to
interesting physics like the spin transfer effect in spin valves. The
dynamics of magnetizations \cite{Igor} is then governed by the parametric
torques due to spin-polarized currents and magnetic fields. This
spin-transfer effect was theoretically proposed \cite{Slonczewski,Berger}
and subsequently verified in experiment \cite{Katine}. Since this novel spin
torque is proposed, many interesting phenomena have been studied, such as
spin wave excitation \cite{Bazaliy,Rezende}, magnetization switching and
reversal \cite{Sun,Wegrowe,Heide,Tsoi02,Chen,Myers}, domain-wall dynamics 
\cite{S. Zhang,Xin}, and magnetic solitons \cite{lizd,pbhe}. In these
studies, the dynamics of magnetization $\mathbf{M}\left( \mathbf{r},t\right) 
$ is described by a modified Landau-Lifshitz equation including the term of
spin-transfer torque. In a typical ferromagnet, the magnetization is rarely
uniform, i.e., the spatial-dependence magnetization, and a new form of spin
torque \cite{Bazaliy,S. Zhang} is proposed in conducting ferromagnetic
structures. With this new form of spin torque, the nonlinear excitations on
the background of ground state are studied, such as the unique features of N%
\'{e}el wall motion in a nanowire \cite{S. Zhang}, kink soliton solution and
the domain wall dynamics in a biaxial ferromagnet \cite{Xin}, bulk spin
excitations \cite{Bazaliy,S. Zhang}, and magnetic soliton solutions for
isotropic case \cite{lizd} and uniaxial anisotropic case \cite{pbhe}. It is
well known that the nonlinear spin wave and magnetic solitons are always
topic research in confined ferromagnetic materials \cite%
{Cottam,Tsoi,Yamada,Belliard} due to the interaction between spin-polarized
current and local magnetization, especially the generation and detection of
magnons excitation \cite{Tsoi} in a magnetic multilayer.

From Eq. (\ref{spincurrent0}) one can obtain the solution of spin current if
magnetization $\mathbf{M}\left( \mathbf{r},t\right) $ is known. For
simpleness we consider an infinite long ferromagnetic nanowire where the
electronic current flows along the long length of the wire, defined as $x$
direction. The $z$ axis is taken as the direction of uniaxial anisotropy
field and the external field. Assuming the magnetization is nonuniform only
along the direction of current, the spin current in Eq. (\ref{spincurrent0})
can be written as 
\begin{equation}
\mathbf{j}\left( x,t\right) =b_{J}\mathbf{M}\left( x,t\right) ,
\label{spincurrent}
\end{equation}%
where $b_{J}=Pj_{e}\mu _{B}/(eM_{s})$. The spatial variation of the spin
current produces a reaction torque on the magnetization as $\tau _{b}\equiv
\partial \mathbf{j}/\partial x=b_{J}\partial \mathbf{M}/\partial x$ which
enter the modified Landau-Lifshitz equation as shown below. As reported in
previous work that spin wave solution and soliton solution \cite{Bazaliy,S.
Zhang,pbhe} are admitted for the magnetization $\mathbf{M}\left( x,t\right) $%
, and then the spin current $\mathbf{j}$ has the periodic form and the
pulsed form, respectively.

In the present paper, we will investigate the properties of spin current $%
\mathbf{j}$ on the background of a periodic solution corresponding to the
soliton solution of magnetization on a nonlinear spin wave background. The
paper is organized as follows. In section II we transform Landau-Lifshitz
equation into an equation of nonlinear Schr\"{o}dinger type in the
long-wavelength approximation. By means of Darboux transformation the
soliton solution for the spin-polarized current are constructed analytically
in section III. In section IV we discuss the properties of the solution for
the spin polarized current in detail. Section V is our conclusion.

\section{Dynamics equation of magnetization in ferromagnetic nanowire}

The dynamics of the localized magnetization is described by the modified
Landau-Lifshitz equation \cite{Bazaliy,S. Zhang} with the term of
spin-transfer torque as 
\begin{equation}
\frac{\partial \mathbf{M}}{\partial t}=-\gamma \mathbf{M\times H}_{\text{eff}%
}+\frac{\alpha }{M_{s}}\mathbf{M\times }\frac{\partial \mathbf{M}}{\partial t%
}+\boldsymbol{\tau }_{b},  \label{LL}
\end{equation}%
where the localized magnetization $\mathbf{M}\equiv \mathbf{M}\left(
x,t\right) $, $\gamma $ is the gyromagnetic ratio, $\alpha $ is the Gilbert
damping parameter, and $\mathbf{H}_{\text{eff}}$ represents the effective
magnetic field including the external field, the anisotropy field, the
demagnetization field and the exchange field. This effective field can be
written as $\mathbf{H}_{\text{eff}}=\left( 2A/M_{s}^{2}\right) \partial ^{2}%
\mathbf{M/}\partial x^{2}\mathbf{+}\left[ \left( H_{K}/M_{s}-4\pi \right)
M_{z}+H_{\text{ext}}\right] \mathbf{e}_{z}$, where $A$ is the exchange
constant, $H_{K}$ is the anisotropy field, $H_{\text{ext}}$ is the applied
external field, and $\mathbf{e}_{z}$ is the unit vector along the $z$
direction. Introducing the normalized magnetization as $\mathbf{m=M/}M_{s}$,
Eq. (\ref{LL}) can be simplified as the dimensionless form 
\begin{eqnarray}
\frac{\partial \mathbf{m}}{\partial t} &=&-(\mathbf{m}\times \frac{\partial
^{2}\mathbf{m}}{\partial x^{2}})+\alpha \mathbf{m\times }\frac{\partial 
\mathbf{m}}{\partial t}+\frac{b_{J}t_{0}}{l_{0}}\frac{\partial \mathbf{m}}{%
\partial x}  \notag \\
&&-(m_{z}+\frac{H_{\text{ext}}}{H_{K}-4\pi M_{s}})(\mathbf{m\times e}_{z}),
\label{LL1}
\end{eqnarray}%
where time $t$ and space coordinate $x$ has been rescaled by the
characteristic time $t_{0}=1/(\gamma (H_{K}-4\pi M_{s}))$ and length $l_{0}=%
\sqrt{2A/((H_{K}-4\pi M_{s})M_{s})}$, respectively.

It is obvious that $\mathbf{m\equiv }\left( m_{x},m_{y},m_{z}\right) =\left(
0,0,1\right) $ forms the ground state of system, and two types of the
nonlinear excited state, i.e., spin wave solution and magnetic soliton, are
admitted for Eq. (\ref{LL1}). When the magnetic field is high enough, the
deviation of magnetization from the ground state is small for two types of
excited state. In this case we can make a reasonable transformation

\begin{equation}
\psi =m_{x}+im_{y},\text{ }m_{z}=\sqrt{1-\left\vert \psi \right\vert ^{2}}.
\label{trans1}
\end{equation}%
Substituting the above equations into Eq. (\ref{LL1}) we obtain%
\begin{eqnarray}
i\frac{\partial \psi }{\partial t} &=&m_{z}\allowbreak \frac{\partial
^{2}\psi }{\partial x^{2}}-\psi \frac{\partial ^{2}m_{z}}{\partial x^{2}}%
-\alpha \left( m_{z}\frac{\partial \psi }{\partial t}-\psi \frac{\partial
m_{z}}{\partial t}\right)  \notag \\
&&+i\frac{b_{J}t_{0}}{l_{0}}\frac{\partial \psi }{\partial x}-\left( m_{z}+%
\frac{H_{\text{ext}}}{H_{K}-4\pi M_{s}}\right) \psi .  \label{nls1}
\end{eqnarray}%
It is easy to get two solutions of Eq. (\ref{nls1}): one is $\psi =0$,
corresponding to the ground state $\mathbf{m}=\left( 0,0,1\right) $, i.e., $%
\mathbf{j}=\left( 0,0,b_{J}M_{s}\right) $, and the other is bulk spin wave
excitations, $\psi =A_{c}e^{i(-k_{c}x+\omega _{c}t)}$, corresponding to the
periodic spin current%
\begin{eqnarray}
j_{x} &=&b_{J}M_{s}A_{c}\cos \left( -k_{c}x\mathbf{+}\omega _{c}t\right) , 
\notag \\
j_{y} &=&b_{J}M_{s}A_{c}\sin \left( -k_{c}x\mathbf{+}\omega _{c}t\right) , 
\notag \\
j_{z} &=&b_{J}M_{s}\sqrt{1-A_{c}^{2}},  \label{magnon}
\end{eqnarray}%
where $k_{c}$ and $\omega _{c}$ are the dimensionless wave number and
frequency of spin wave, and the transverse amplitude $A_{c}<<1$. For the
attractive interaction the nonlinear spin waves in ferromagnet with
anisotropy lead to the macroscopic phenomena, i.e., the appearance of
spatially localized magnetic excited state (magnetic soliton).

In the present paper we want to obtain the soliton solution of magnetization
on a nonlinear spin wave background in a uniaxial ferromagnetic nanowire
with spin torque. However, Eq. (\ref{nls1}) is not integrable. To our
purpose we consider the case of without damping and the long-wavelength
approximation \cite{Kosevich}, where the dimensionless wave number $k_{c}\ll
1$. Keeping only the nonlinear terms of the order of the magnitude of $%
\left\vert \psi \right\vert ^{2}\psi $, Eq. (\ref{nls1}) can be simplified
as an integrable equation%
\begin{equation}
i\frac{\partial \psi }{\partial t}\!=\!\frac{\partial ^{2}\psi }{\partial
x^{2}}\!+\!\frac{1}{2}\left\vert \psi \right\vert ^{2}\psi \!-\!(1+\frac{H_{%
\text{ext}}}{H_{K}\!-\!4\pi M_{s}})\psi \!+\!i\frac{b_{J}t_{0}}{l_{0}}\frac{%
\partial \psi }{\partial x},  \label{nls2}
\end{equation}%
whose soliton solutions on the background of the ground state, $\psi =0$,
can be obtained by Hirota methods \cite{pbhe,Hirota}. In order to discuss
the properties of soliton solution on the background of spin wave, here we
use a straightforward Darboux transformation \cite{Matveev,Gu,Lilu} to
construct general expressions of soliton solution of Eq. (\ref{nls2}) with
which the soliton solution for the spin-polarized current are obtained from
Eqs. (\ref{spincurrent}) and (\ref{trans1}). For this reason we will
consider mainly the solutions of Eq. (\ref{nls2}) in the following section.

The main idea of Darboux transformation is that it firstly transforms the
nonlinear equation into the Lax representation, and then in terms of a
series of transformations the soliton solution can be constructed
algebraically with an obvious seed solution of the nonlinear equation. In
terms of Ablowitz-Kaup-Newell-Segur technique Lax representation for Eq. (%
\ref{nls2}) can be constructed as 
\begin{eqnarray}
\frac{\partial \Psi }{\partial x} &=&U\Psi ,  \notag \\
\frac{\partial \Psi }{\partial t} &=&V\Psi ,  \label{laxE}
\end{eqnarray}%
where $\Psi =\left( 
\begin{array}{cc}
\Psi _{1} & \Psi _{2}%
\end{array}%
\right) ^{T}$, the superscript \textquotedblleft $T$\textquotedblright\
denotes the matrix transpose, and the Lax pairs $U$ and $V$ are defined by%
\begin{eqnarray}
U &=&\lambda \sigma _{3}+q,  \notag \\
V &=&\left( -i2\lambda ^{2}+\lambda \alpha _{1}+\alpha _{2}\right) \sigma
_{3}  \notag \\
&&+\left( \alpha _{1}-i2\lambda \right) q+i\left( \frac{\partial q}{\partial
x}+q^{2}\right) \sigma _{3},  \label{laxpair}
\end{eqnarray}%
where 
\begin{eqnarray*}
\alpha _{1} &=&\frac{b_{J}t_{0}}{l_{0}},\alpha _{2}=i\frac{1}{2}\left( 1+%
\frac{H_{\text{ext}}}{H_{K}-4\pi M_{s}}\right) , \\
\sigma _{3} &=&\left( 
\begin{array}{ll}
1 & 0 \\ 
0 & -1%
\end{array}%
\right) ,\text{ }q=\frac{1}{2}\left( 
\begin{array}{ll}
0 & \psi \\ 
-\overline{\psi } & 0%
\end{array}%
\right) ,
\end{eqnarray*}%
where $\lambda $ is the complex spectral parameter, and the overbar denotes
the complex conjugate. With the natural condition of Eq. (\ref{laxE}) $%
\partial ^{2}\Psi /\left( \partial x\partial t\right) =\partial ^{2}\Psi
/\left( \partial t\partial x\right) $, i.e., $\partial U/\partial t-\partial
V/\partial x+\left[ U,V\right] =0$ the integrable Eq. (\ref{nls2}) can be
recovered successfully. Now the Eqs. (\ref{laxE}) and (\ref{laxpair}) have
made the normal form of the developed Darboux transformation with which we
can get the general N-soliton solution as shown in next section.

\section{Darboux transformation}

In this section we briefly introduce the procedure for getting soliton
solution for the developed Darboux transformation. Consider the following
transformation 
\begin{equation}
\Psi \left[ 1\right] =\left( \lambda I-K\right) \Psi ,\text{ }
\label{Darboux1}
\end{equation}%
where $K=S\Lambda S^{-1}$, $\Lambda =$diag$\left( \lambda _{1},\lambda
_{2}\right) $, and $S$ \hspace{0in}is a nonsingular matrix which satisfies 
\begin{equation}
\frac{\partial }{\partial x}S=\sigma _{3}S\Lambda +qS.  \label{Darboux2}
\end{equation}%
Letting $\Psi \left[ 1\right] $ satisfy the Lax equation 
\begin{equation}
\frac{\partial }{\partial x}\Psi \left[ 1\right] =U_{1}\Psi \left[ 1\right] ,
\label{Darboux3}
\end{equation}%
where%
\begin{eqnarray*}
U_{1} &=&\lambda \sigma _{3}+q_{1}, \\
q_{1} &=&\frac{1}{2}\left( 
\begin{array}{ll}
0 & \psi _{1} \\ 
-\overline{\psi }_{1} & \text{ \hspace{0in} \hspace{0in} }0%
\end{array}%
\right) .
\end{eqnarray*}%
With the help of Eqs. (\ref{laxpair}), (\ref{Darboux1}) and (\ref{Darboux2}%
), we obtain the Darboux transformation from Eq. (\ref{Darboux3}) in the form%
\begin{equation}
\psi _{1}=\psi +4K_{12},  \label{Darboux4}
\end{equation}%
which shows that a new solution $\psi _{1}$ of Eq. (\ref{nls2}) with
\textquotedblleft seed\textquotedblright\ solution $\psi $ can be obtained
if $K$ is known.

It is verified easily that, if $\Psi =\left( 
\begin{array}{cc}
\Psi _{1} & \Psi _{2}%
\end{array}%
\right) ^{T}$ is the eigenfunction of Eq. (\ref{laxE}) with eigenvalue $%
\lambda =\lambda _{1}$, then $\left( 
\begin{array}{cc}
-\overline{\Psi }_{2} & \overline{\Psi }_{1}%
\end{array}%
\right) ^{T}$ is also the eigenfunction, however, with eigenvalue $-%
\overline{\lambda }_{1}$. Therefore $S$ and $\Lambda $ can be taken the form 
$\allowbreak $ $\allowbreak $ 
\begin{equation}
S=\left( 
\begin{array}{ll}
\Psi _{1} & -\overline{\Psi }_{2} \\ 
\Psi _{2} & \text{ }\overline{\Psi }_{1}%
\end{array}%
\right) ,\Lambda =\left( 
\begin{array}{ll}
\lambda _{1} & \text{ }0 \\ 
0 & -\overline{\lambda }_{1}%
\end{array}%
\right) ,  \label{Darboux5}
\end{equation}%
which ensures that Eq. (\ref{Darboux2}) is held. Then Eq. (\ref{Darboux4})
becomes 
\begin{equation}
\psi _{1}=\psi +4\left( \lambda _{1}+\overline{\lambda }_{1}\right) \frac{%
\Psi _{1}\overline{\Psi }_{2}}{\Psi ^{T}\overline{\Psi }},  \label{Darboux}
\end{equation}%
where $\Psi ^{T}\overline{\Psi }=\left\vert \Psi _{1}\right\vert
^{2}+\left\vert \Psi _{2}\right\vert ^{2}$, $\Psi =\left( \Psi _{1},\Psi
_{2}\right) ^{T}$ to be determined is the eigenfunction of Eq. (\ref{laxE})
corresponding to the eigenvalue $\lambda _{1}$ for the solution $\psi $ of
Eq. (\ref{nls2}). Thus by solving Eq. (\ref{laxE}) we can generate a new
solution for $\mathbf{j}$ with the help of Eqs. (\ref{spincurrent}), (\ref%
{trans1}) and (\ref{nls2}) from an obvious \textquotedblleft
seed\textquotedblright\ solution of Eq. (\ref{nls2}).

To obtain exact $N$-order solution, we firstly rewrite the Darboux
transformation in Eq. (\ref{Darboux}) as in the form 
\begin{equation}
\psi _{1}=\psi +4\left( \lambda _{1}+\overline{\lambda }_{1}\right) \frac{%
\Psi _{1}\left[ 1,\lambda _{1}\right] \overline{\Psi }_{2}\left[ 1,\lambda
_{1}\right] }{\Psi \left[ 1,\lambda _{1}\right] ^{T}\overline{\Psi }\left[
1,\lambda _{1}\right] },  \label{Oneso}
\end{equation}%
where $\Psi \left[ 1,\lambda \right] =\left( \Psi _{1}\left[ 1,\lambda %
\right] ,\Psi _{2}\left[ 1,\lambda \right] \right) ^{T}$ denotes the
eigenfunction of Eq. (\ref{laxE}) corresponding to eigenvalue $\lambda $.
Then repeating above the procedure $N$ times, we can obtain the exact $N$%
-order solution 
\begin{equation}
\psi _{N}=\psi +4\sum_{n=1}^{N}(\lambda _{n}+\overline{\lambda }_{n})\frac{%
\Psi _{1}[n,\lambda _{n}]\overline{\Psi }_{2}[n,\lambda _{n}]}{\Psi \lbrack
n,\lambda _{n}]^{T}\overline{\Psi }[n,\lambda _{n}]},  \label{Multiso}
\end{equation}%
where 
\begin{align*}
\Psi \left[ n,\lambda \right] & =\left( \lambda -K\left[ n-1\right] \right)
\cdots \left( \lambda -K\left[ 1\right] \right) \Psi \left[ 1,\lambda \right]
, \\
K_{l_{1}l_{2}}[n^{\prime }]& =(\lambda _{n^{\prime }}+\overline{\lambda }%
_{n^{\prime }})\frac{\Psi _{l_{1}}[n^{\prime },\lambda _{n^{\prime }}]%
\overline{\Psi }_{l_{2}}[n^{\prime },\lambda _{n^{\prime }}]}{\Psi \lbrack
n^{\prime },\lambda _{n^{\prime }}]^{T}\overline{\Psi }[n^{\prime },\lambda
_{n^{\prime }}]}-\overline{\lambda }_{n^{\prime }}\delta _{l_{1}l_{2}},
\end{align*}%
here $\Psi \left[ n^{\prime },\lambda \right] $ is the eigenfunction
corresponding to $\lambda _{n^{\prime }}$ for $\Psi _{n^{\prime }-1}$ with $%
\Psi _{0}\equiv \Psi $, and $l_{1},l_{2}=1,2$, $n^{\prime }=1,2,\cdots ,n-1$%
, $n=2,3,\cdots ,N$. Thus if choosing a \textquotedblleft
seed\textquotedblright\ as the basic initial solution, by solving linear
characteristic equation system (\ref{laxE}), one can construct a set of new
solutions from Eq. (\ref{Multiso}).

In the following we take the initial \textquotedblleft
seed\textquotedblright\ solution $\psi =A_{c}e^{i(-k_{c}x\mathbf{+}\omega
_{c}t)}$ corresponding to a periodic solution (\ref{magnon}), where the
dispersion relation, $\omega
_{c}=k_{c}^{2}-k_{c}b_{J}t_{0}/l_{0}-A_{c}^{2}/2+\left[ 1+H_{ext}/\left(
H_{K}-4\pi M_{s}\right) \right] $, is obtained from Eq. (\ref{nls2}). After
the tedious calculation for solving the linear equation system (\ref{laxE})
we have the eigenfunction corresponding to eigenvalue $\lambda $ in the form 
$\allowbreak $ $\allowbreak $ $\allowbreak $ $\allowbreak $ $\allowbreak $

\begin{align}
\Psi _{1}& =LC_{1}e^{\Theta _{1}}+\frac{1}{2}A_{c}C_{2}e^{\Theta _{2}}, 
\notag \\
\Psi _{2}& =\frac{1}{2}A_{c}C_{1}e^{-\Theta _{2}}+LC_{2}e^{-\Theta _{1}},
\label{Laxsolution}
\end{align}%
$\allowbreak \allowbreak $where the parameters $C_{1}$ and $C_{2}$ are the
arbitrary complex constants, and the other parameters are defined by%
\begin{eqnarray}
\Theta _{1} &=&-\frac{1}{2}i\left( k_{c}x-\omega _{c}t\right) +D\left(
x+\delta t\right) ,  \notag \\
\Theta _{2} &=&-\frac{1}{2}i\left( k_{c}x-\omega _{c}t\right) -D\left(
x+\delta t\right) ,  \notag \\
L &=&-\frac{1}{2}ik_{c}-D-\lambda ,  \notag \\
D &=&\frac{1}{2}\sqrt{\left( ik_{c}+2\lambda \right) ^{2}-A_{c}^{2}},  \notag
\\
\delta &=&-i2\lambda -k_{c}+\frac{b_{J}t_{0}}{l_{0}}.  \label{para}
\end{eqnarray}%
With the help of the formulas (\ref{spincurrent}), (\ref{trans1}), (\ref%
{Multiso}), and (\ref{Laxsolution}) we can obtain the desired soliton
solution for the spin current.

\section{Properties of soliton solution for the spin-polarized current}

Taking the spectral parameter $\lambda =\lambda _{1}\equiv \mu _{1}/2+i\nu
_{1}/2$, here $\mu _{1}$ and $\nu _{1}$ are real number, in Eq. (\ref%
{Laxsolution}), and substituting them into Eqs. (\ref{Oneso}) and (\ref%
{trans1}) we obtain the one-soliton solution for the spin-polarized current
from Eq. (\ref{spincurrent}) as 
\begin{eqnarray}
j_{x} &=&b_{J}M_{s}[A_{c}\cos \varphi +\frac{2\mu _{1}}{\Delta _{1}}\left(
Q_{1}\cos \varphi -Q_{2}\sin \varphi \right) ],  \notag \\
j_{y} &=&b_{J}M_{s}[A_{c}\sin \varphi +\frac{2\mu _{1}}{\Delta _{1}}\left(
Q_{1}\sin \varphi +Q_{2}\cos \varphi \right) ],  \notag \\
j_{z} &=&b_{J}M_{s}\sqrt{1-(A_{c}+\frac{2\mu _{1}Q_{1}}{\Delta _{1}})^{2}-(%
\frac{2\mu _{1}Q_{2}}{\Delta _{1}})^{2}},  \label{onesoliton}
\end{eqnarray}%
where 
\begin{eqnarray}
\theta _{1} &=&2D_{1R}x+2\left( D_{1}\delta _{1}\right) _{R}t+2x_{0},  \notag
\\
\Phi _{1} &=&2D_{1I}x+2\left( D_{1}\delta _{1}\right) _{I}t-2\varphi _{0}, 
\notag \\
\varphi &=&-k_{c}x+\omega _{c}t,  \label{para1}
\end{eqnarray}%
\begin{eqnarray}
Q_{1} &=&A_{c}L_{1R}\cosh \theta _{1}+\left( \left\vert L_{1}\right\vert
^{2}+\frac{1}{4}A_{c}^{2}\right) \cos \Phi _{1},  \notag \\
Q_{2} &=&A_{c}L_{1I}\sinh \theta _{1}+\left( \left\vert L_{1}\right\vert
^{2}-\frac{1}{4}A_{c}^{2}\right) \sin \Phi _{1},  \notag \\
\Delta _{1} &=&\left( \left\vert L_{1}\right\vert ^{2}+\frac{1}{4}%
A_{c}^{2}\right) \cosh \theta _{1}+A_{c}L_{1R}\cos \Phi _{1},  \notag
\end{eqnarray}%
where the subscript $R$ and $I$ represent the real part and imaginary part,
respectively. The other parameters are $D_{1}=\sqrt{\left( ik_{c}/2+\lambda
_{1}\right) ^{2}-A_{c}^{2}/4}$, $L_{1}=-ik_{c}/2-D_{1}-\lambda _{1}$, $%
\delta _{1}=-i2\lambda _{1}-k_{c}+b_{J}t_{0}/l_{0}$, $x_{0}=-\left( \ln
\left\vert C_{2}/C_{1}\right\vert \right) /2$, and $\varphi _{0}=\left[ \arg
\left( C_{2}/C_{1}\right) \right] /2$, where $C_{1},C_{2}$ are the arbitrary
complex constants.

The solution (\ref{onesoliton}) describes a one-soliton solution for the
spin-polarized current in ferromagnetic nanowire embedded in the periodic
spin current background (\ref{magnon}): (a) When $\mu _{1}=0$, the solution (%
\ref{onesoliton}) reduces to the periodic solution (\ref{magnon}). (b) When
the spin-wave amplitude vanishes, namely $A_{c}=0$, the solution (\ref%
{onesoliton}) reduces to the solution in the form%
\begin{eqnarray}
j_{x} &=&\frac{2\mu _{1}b_{J}M_{s}}{\cosh \theta _{1}}\cos \left( \Phi
_{1}+\eta \right) ,  \notag \\
j_{y} &=&\frac{2\mu _{1}b_{J}M_{s}}{\cosh \theta _{1}}\sin \left( \Phi
_{1}+\eta \right) ,  \notag \\
j_{z} &=&b_{J}M_{s}\sqrt{1-\frac{4\mu _{1}^{2}}{\cosh ^{2}\theta _{1}}},
\label{onesoliton1}
\end{eqnarray}%
where 
\begin{eqnarray}
\theta _{1} &=&\mu _{1}[x+(2\nu _{1}+\frac{b_{J}t_{0}}{l_{0}})t+\frac{2}{\mu
_{1}}x_{0}],  \notag \\
\Phi _{1} &=&\nu _{1}\{x-[\frac{1}{\nu _{1}}\left( \mu _{1}^{2}-\nu
_{1}^{2}\right) -\allowbreak \frac{b_{J}t_{0}}{l_{0}}]t-\frac{2}{\nu _{1}}%
\varphi _{0}\},  \notag \\
\eta &=&(1+\frac{H_{ext}}{H_{K}-4\pi M_{s}})t.  \label{para2}
\end{eqnarray}%
The solution (\ref{onesoliton1}) is in fact the same as the solution (8) in
Ref. \cite{pbhe} (One should notice that the transformation (\ref{trans1})
is different from that in Ref. \cite{pbhe}).

The solution (\ref{onesoliton1}) indicates the spatially localized
excitation \cite{wall} which is denoted by the transverse amplitude $2\mu
_{1}$ deviated from the ground state $\mathbf{j}=\left(
0,0,b_{J}M_{s}\right) $. The components $j_{x}$ and $j_{y}$ precess around
the component $j_{z}$ with the frequency $\Omega _{1}=\nu _{1}^{2}-\mu
_{1}^{2}+b_{J}t_{0}/l_{0}+1+H_{ext}/\left( H_{K}-4\pi M_{s}\right) $, and
the soliton solution is characterized by the width $1/\mu _{1}$, the
velocity of soliton center $v_{1}=-\left( 2\nu _{1}+b_{J}t_{0}/l_{0}\right) $%
. The wave number $k_{s,1}=-\nu _{1}$ and the frequency $\Omega _{1}$ of the
\textquotedblleft carrier wave\textquotedblright\ are related by the
dispersion law $\Omega _{1}=k_{s,1}^{2}-\allowbreak
k_{s,1}b_{J}t_{0}/l_{0}-\mu _{1}^{2}+1+H_{ext}/\left( H_{K}-4\pi
M_{s}\right) $ which shows that the magnetic field contribute to precession
frequency only. The magnetic soliton energy is seen to be $%
E_{1}=-b_{J}^{2}t_{0}^{2}/\left( 2l_{0}\right) ^{2}-\mu
_{1}^{2}+1+H_{ext}/\left( H_{K}-4\pi M_{s}\right) $ $+\frac{1}{2}m^{\ast
}v_{1}^{2}$, where the dimensionless effective mass $m^{\ast }$ of soliton
is $1/2$.

From Eqs. (\ref{onesoliton1}) and (\ref{para2}) we also see that the term $%
b_{J}$ can change the velocity and the precessional frequency of soliton on
a background of\ the ground state $\mathbf{j}=\left( 0,0,b_{J}M_{s}\right) $%
. This case confirms the previous study \cite{S. Zhang,pbhe}. However, on
the background (\ref{magnon}) novel properties of Eq. (\ref{onesoliton})
will be described below. The properties of envelope soliton solution (\ref%
{onesoliton}) are characterized by the width $1/\left( 2D_{1R}\right) $, the
wave number $k_{s}=-2D_{1I}$, the initial center position $-x_{0}/D_{1R}$,
and the envelope velocity $v_{1}=-\left( D_{1}\delta _{1}\right) _{R}/D_{1R}$%
. The initial center position of soliton is moved $x_{0}\left( 2/\mu
_{1}-1/D_{1R}\right) $ by the spin wave which show new way controlling the
soliton in space.

From the expressions of $D_{1}$\ and $\delta _{1}$\ we find that the
velocity and the width of envelope soliton are modulated by the amplitude $%
A_{c}$\ and wave number $k_{c}$\ of spin wave as shown in Figure 1. From
figure 1(a) and 1(c) we see that absolute value of velocity and the width of
envelope soliton become large with the increasing $A_{c}$. Figure 1(b) shows
that the value of $k_{c}$ nearby $-\nu _{1}$ has obvious effect on the
velocity of soliton. When $k_{c}=-\nu _{1}$, the width of soliton is maximal
as shown in Figure 1(d).

From Eq. (\ref{para1}) we can directly see that when $D_{1I}\delta
_{1I}=\delta _{1R}D_{1R}$, the parameters $\theta _{1}$ depends only on $x$
which implies the envelope velocity $-\left( D_{1}\delta _{1}\right)
_{R}/D_{1R}$ becomes zero, i.e., the soliton is trapped in space by the
nonlinear spin wave. It should be noted that this condition can be written
as 
\begin{equation}
j_{e}=\frac{eM_{s}}{P\mu _{B}}\frac{l_{0}}{t_{0}}\left( -\mu _{1}\frac{D_{1I}%
}{D_{1R}}-\nu _{1}+k_{c}\right) ,  \label{trap1}
\end{equation}%
which is determined by the characteristic velocity $l_{0}/t_{0}$, the
amplitude and wave number of soliton and the nonlinear spin-wave, and the
parameters $\left( eM_{s}/P\mu _{B}\right) $, respectively. From a tedious
calculation we found that when $A_{c}<<k_{c}$ and $\mu _{1}<<\nu _{1}$ the
condition in Eq. (\ref{trap1}) reduces to $j_{e}\approx -2\nu _{1}\left(
l_{0}/t_{0}\right) \left( eM_{s}/P\mu _{B}\right) $. When the amplitude of
spin wave vanishes, namely $A_{c}=0$, the trapping condition in Eq. (\ref%
{trap1}) reduces to $j_{e}=-2\nu _{1}\left( l_{0}/t_{0}\right) \left(
eM_{s}/P\mu _{B}\right) $ which is determined by the characteristic velocity 
$l_{0}/t_{0}$, the soliton wave number $\nu _{1}$, and the parameters $%
\left( eM_{s}/P\mu _{B}\right) $. These results show that the background has
almost no effect on the trapping condition in the special case $A_{c}<<k_{c}$
and $\mu _{1}<<\nu _{1}$. For the materials of Co$_{3}$Pt alloy films \cite%
{Yamada} which has high perpendicular anisotropy, we chose $H_{K}=1\times
10^{4}$Oe, $A=1.0\times 10^{-6}$erg$/$cm, $4\pi M_{s}=1\times 10^{2}$Oe, $%
\gamma =1.76\times 10^{7}$\ Oe$^{-1}$s$^{-1}$, $P=0.35$, and the
dimensionless parameters $k_{c}=0.05,A_{c}=0.02,\nu _{1}=-0.12,$\ and $\mu
_{1}=0.1$. The critical electric current trapping soliton is $%
j_{e}=1.\,\allowbreak 867\times 10^{4}$A/cm$^{2}$\textbf{.} It is very
important to point out that from Eqs. (\ref{onesoliton}), (\ref{para1}), and
expressions of $\delta _{1}$ the term $b_{J}$ change not only the velocity,
but also the frequency effecting on soliton energy . This property trapping
the soliton in space by the nonlinear spin wave is characterized by the
spatial and temporal period along the direction of soliton propagation, $%
x=-\left( D_{1}\delta _{1}\right) _{R}t/D_{1R}-x_{0}/D_{1R}$, denoted by $%
\pi \left( D_{1}\delta _{1}\right) _{R}/\left[ \delta _{1I}\left(
D_{1R}^{2}+D_{1I}^{2}\right) \right] $ and $\pi D_{1R}/\left[ \delta
_{1I}\left( D_{1R}^{2}+D_{1I}^{2}\right) \right] $, respectively.

In order to explain some novel properties of solution (\ref{onesoliton}) we
discuss the special case $k_{c}=-\nu _{1}$ and analyze two representative
situations in detail: (a) The amplitude $A_{c}$ exceeds the half of the
transverse amplitude $2\mu _{1}$ of soliton. (b) The amplitude $A_{c}$ is
less than the half of transverse amplitude $2\mu _{1}$ of soliton (which
implies $A_{c},\mu _{1}>0$).

(a) In the case $\mu _{1}^{2}<A_{c}^{2}$ the solution (\ref{onesoliton})
reduces to 
\begin{eqnarray}
j_{x} &\!=\!&b_{J}M_{s}\left( R_{1}^{\prime }\cos \varphi -R_{2}^{\prime
}\sin \varphi \right) ,  \notag \\
j_{y} &\!=\!&b_{J}M_{s}\left( R_{1}^{\prime }\sin \varphi +R_{2}^{\prime
}\cos \varphi \right) ,  \notag \\
j_{z} &\!=\!&b_{J}M_{s}\sqrt{1\!-\!A_{c}^{2}\!-\!\frac{\zeta _{1}(A_{c}\cosh
\theta _{1}\cos \Phi _{1}-\mu _{1})}{(A_{c}\cosh \theta _{1}-\mu _{1}\cos
\Phi _{1})^{2}}},  \label{onesoliton2}
\end{eqnarray}%
where $\varphi $ is given in Eq. (\ref{para1}), and the other parameters are
determined by%
\begin{eqnarray*}
\zeta _{1} &=&4\mu _{1}\kappa _{1}^{2}, \\
\kappa _{1} &=&\sqrt{A_{c}^{2}-\mu _{1}^{2}}, \\
R_{1}^{\prime } &=&-A_{c}+\frac{2\kappa _{1}^{2}\cosh \theta _{1}}{%
A_{c}\cosh \theta _{1}-\mu _{1}\cos \Phi _{1}}, \\
R_{2}^{\prime } &=&-\frac{2\mu _{1}\kappa _{1}\sinh \theta _{1}}{A_{c}\cosh
\theta _{1}-\mu _{1}\cos \Phi _{1}},
\end{eqnarray*}%
\begin{eqnarray}
\theta _{1} &=&\mu _{1}\kappa _{1}t+2x_{0},  \notag \\
\Phi _{1} &=&\kappa _{1}\left( x-v_{1}t-2\varphi _{0}/\kappa _{1}\right) .
\label{para3}
\end{eqnarray}%
A simple analysis for Eq. (\ref{para3}) reveals that the solution (\ref%
{onesoliton2}) is periodic in the space coordinate, denoted by $2\pi /\kappa
_{1}$, and aperiodic in the temporal variable, as shown in Fig. 2. From Fig.
2 we can see that the background becomes unstable, therefore the solution (%
\ref{onesoliton2}) can be considered as describing the modulation
instability process \cite{Ablowitz}. Along the propagation direction of
soliton the expression of $j_{z}$ has a maximum $j_{z}=b_{J}M_{s}$, i.e., $%
m_{z}=1$, at $\cos \Phi _{1}=\left( 2\mu _{1}^{2}-A_{c}^{2}\right) /\mu
_{1}A_{c}$ when $A_{c}^{2}/4<\mu _{1}^{2}<A_{c}^{2}$, which shows these
points are not excited even on the spin wave background, and has a minimum $%
j_{z}=b_{J}M_{s}[1-\left( 2\mu _{1}+A_{c}\right) ^{2}]^{1/2}$ at $\sin \Phi
_{1}=0$. When $\mu _{1}^{2}<A_{c}^{2}/4$, the expression of $j_{z}$ has a
maximum $j_{z}=b_{J}M_{s}[1-\left( 2\mu _{1}-A_{c}\right) ^{2}]^{1/2}$ at $%
\Phi _{1}=0$, and a minimum $j_{z}=b_{J}M_{s}[1-\left( 2\mu
_{1}+A_{c}\right) ^{2}]^{1/2}$ at $\Phi _{1}=\pi $. These results show that
the linear combined transverse amplitude of spin wave and magnetic soliton
can be obtained in these special cases.

(b) In the case $\mu _{1}^{2}>A_{c}^{2}$ the solution (\ref{onesoliton})
reduces to%
\begin{eqnarray}
j_{x} &\!=\!&b_{J}M_{s}\left( R_{1}\cos \varphi -R_{2}\sin \varphi \right) ,
\notag \\
j_{y} &\!=\!&b_{J}M_{s}\left( R_{1}\sin \varphi +R_{2}\cos \varphi \right) ,
\notag \\
j_{z} &\!=\!&b_{J}M_{s}\sqrt{1\!-\!A_{c}^{2}\!-\!\frac{\zeta _{2}(\mu
_{1}\!-\!A_{c}\cosh \theta _{1}\cos \Phi _{1})}{(\mu _{1}\cosh \theta
_{1}\!-\!A_{c}\cos \Phi _{1})^{2}}},  \label{onesoliton3}
\end{eqnarray}%
$\allowbreak \allowbreak $ $\allowbreak $ $\allowbreak $where $\allowbreak $%
\begin{eqnarray}
\zeta _{2} &=&4\mu _{1}\kappa _{2}^{2},  \notag \\
\kappa _{2} &=&\sqrt{\mu _{1}^{2}-A_{c}^{2}}, \\
R_{1} &=&-A_{c}+\frac{2\kappa _{2}^{2}\cos \Phi _{1}}{\mu _{1}\cosh \theta
_{1}-A_{c}\cos \Phi _{1}},  \notag \\
R_{2} &=&\frac{2\mu _{1}\kappa _{2}\sin \Phi _{1}}{\mu _{1}\cosh \theta
_{1}-A_{c}\cos \Phi _{1}},  \notag \\
\theta _{1} &=&\kappa _{2}\left( x-v_{1}t+2x_{0}/\kappa _{2}\right) ,  \notag
\\
\Phi _{1} &=&-\mu _{1}\kappa _{2}t-2\varphi _{0}.  \label{para4}
\end{eqnarray}%
With the expressions (\ref{onesoliton3}) and (\ref{para4}) we can see the
main characteristic of soliton solution: (1) The soliton has the same
envelope velocity $v_{1}=-\left( 2\nu _{1}+b_{J}t_{0}/l_{0}\right) $ on both
the background of a periodic spin current in Eq. (\ref{magnon}) and the
ground state background $\mathbf{j}=\left( 0,0,b_{J}M_{s}\right) $. (2) The
amplitude of $j_{z}$ in Eq. (\ref{onesoliton3}) has the temporal periodic
oscillation as shown in Figure 3. A detail calculation shows that the
amplitude of $j_{z}$ in Eq. (\ref{onesoliton3}) has a minimum at $\theta
_{1}=0$, which is given by$\allowbreak \allowbreak $%
\begin{equation}
j_{z}=b_{J}M_{s}\sqrt{1-A_{c}^{2}-\frac{4\mu _{1}\left( \mu
_{1}^{2}-A_{c}^{2}\right) }{\mu _{1}-A_{c}\cos \Phi _{1}}},  \label{min}
\end{equation}%
and has a maximum%
\begin{equation}
j_{z}=b_{J}M_{s}\sqrt{1-\frac{\mu _{1}^{2}A_{c}^{2}\sin ^{2}\Phi _{1}}{\mu
_{1}^{2}-A_{c}^{2}\cos ^{2}\Phi _{1}}},  \label{max}
\end{equation}%
at $\cosh \theta _{1}=2\mu _{1}/\left( A_{c}\cos \Phi _{1}\right) -\left(
A_{c}\cos \Phi _{1}\right) /\mu _{1}$. Fig. 4(a) presents the evolution
along the propagation direction of the minimum and maximum intensities given
by Eqs. (\ref{min}) and (\ref{max}) (see, respectively, the dashed and
dotted lines), and the spin wave intensity (solid line). The location of
minimum and maximum amplitude (solid and dotted lines, respectively) in the
time-space plane is shown in Fig. 4(b). From Fig. 4, it is seen that the
narrower the soliton, the sharper the peak and the deeper the two dips at
the wings of the soliton. This feature illustrates the characteristic
breather behavior of the soliton as it propagates on the background of a
periodic solution for the spin current in ferromagnetic nanowire.

\section{Conclusion}

In summary, by transforming the modified Landau-Lifshitz equation into an
equation of nonlinear Schr\"{o}dinger type, we study the interaction of a
periodic solution and one-soliton solution for the spin-polarized current in
a uniaxial ferromagnetic nanowire. Our results show that the amplitude of
soliton solution has the spatial and temporal period on the background of a
periodic spin current. The effective mass of soliton is obtained. Moreover,
we found that the soliton can be trapped only in space. We also analyze the
modulation instability and dark soliton on the background of a periodic spin
current which shows the characteristic breather behavior of the soliton as
it propagates along the ferromagnetic nanowire.

\section{Acknowledgment}

This work was supported by NSF of China under grant 90403034, 90406017,
60525417, 10647122, NKBRSF of China under 2005CB724508, 2006CB921400, the
Natural Science Foundation of Hebei Province of China No. A2007000006, the
Foundation of Education Bureau of Hebei Province of China No. 2006110, and
the key subject construction project of Hebei Provincial University of China.

\textbf{Figure Captions}\newline

Fig. 1. (Color online) Velocity and width variety of soliton solution for
spin-polarized current vs the amplitude $A_{c}$ and wave number $k_{c}$ of
spin wave, respectively. The parameters are $\mu _{1}=0.1$, $l_{0}=2\times
10^{-8}$cm, $t_{0}=5.7392\times 10^{-12}$s, and $b_{J}=52$cm/s. (1a)
Velocity vs the amplitude $A_{c}$, $k_{c}=0.1$, $\nu _{1}=-0.12$ (red line); 
$k_{c}=-0.1$, $\nu _{1}=0.12$ (blue dotted line). (1b) Velocity vs spin wave
number $k_{c}$, $A_{c}=0.06$, and $\nu _{1}=-0.15$. (1c) Width vs the
amplitude $A_{c}$, $k_{c}=-0.1$, $\nu _{1}=0.12$. (1d) Width vs spin wave
number $k_{c}$, $A_{c}=0.06$, and $\nu _{1}=-0.15$.

Fig. 2. The evolution of solution (\ref{onesoliton2}) with condition $%
k_{c}=-\nu _{1}$, $\mu _{1}^{2}<A_{c}^{2}$, and the parameters are $\mu
_{1}=0.12,\nu _{1}=-0.1,A_{c}=0.16,b_{J}=34.8$cm/s, $l_{0}=2\times 10^{-10}$%
m, $t_{0}=5.7392\times 10^{-12}$s, $x_{0}=-1.27$, and $\varphi _{0}=0$. The
spin current $j_{z}$ is in unit of $b_{J}M_{s}$, and the same as in Fig. 3
and Fig. 4.

Fig. 3. The evolution of solution (\ref{onesoliton3}) with the condition $%
k_{c}=-\nu _{1}$ and $\mu _{1}^{2}>A_{c}^{2}$. The parameters are $\mu
_{1}=0.1,\nu _{1}=-0.08,A_{c}=0.06,b_{J}=41.8$cm/s, $l_{0}=2\times 10^{-10}$%
m, $t_{0}=5.7392\times 10^{-12}$s, $x_{0}=3$, and $\varphi _{0}=0$.

Fig. 4. (a) The evolution of the minimum amplitude of $j_{z}$ in Eq. (\ref%
{min}) (dashed line), the maximum amplitude of $j_{z}$ in Eq. (\ref{max})
(dotted line), and the background amplitude of $j_{z}$ (solid line) in Eq. (%
\ref{magnon}); (b) The location of the minimum amplitude (solid line) and
the maximum amplitude (dotted line) in the time-propagation distance plane.
The parameters are the same in Fig. 3.

\end{document}